\documentclass{IEEEtran} 
\usepackage{graphicx,url}
\usepackage[T1]{fontenc}
\usepackage[utf8]{inputenc}

\newtheorem{ppat}{Privacy Pattern}

\begin{document}

\title{Privacy Patterns}
\author{Clark Thomborson}

\maketitle

\begin{abstract}
  Inspired by the design patterns of object-oriented software
  architecture, we offer an initial set of ``privacy patterns''.  Our
  intent is to describe the most important ways in which software
  systems can offer privacy to their stakeholders.  We express our
  privacy patterns as class diagrams in the UML (Universal Modelling
  Language), because this is a commonly-used language for expressing
  the high-level architecture of an object-oriented system.  In this
  initial set of privacy patterns, we sketch how each of Westin's four
  states of privacy can be implemented in a software system.  In
  addition to Westin's states of Solitude, Intimacy, Anonymity, and
  Reserve, we develop a privacy pattern for an institutionalised form
  of Intimacy which we call Confidence.
\end{abstract}


\section{Introduction}

Privacy requirements are problematic for any globally-accessible
computer system, because of significant differences in the conceptions
of privacy by stakeholders with various legal traditions, cultures,
religions, and individual desires.  We do not attempt to survey these
variations here.  Instead, we elicit an initial set of ``privacy
patterns'' from Westin's influential monograph \cite{Westin67}, in
which he identified four primary ``states'' of privacy: Solitude,
Intimacy, Anonymity, and Reserve.

In this article, we provide initial answers to the following questions.
\begin{itemize}
\item What architectural features of a software system will make it
  possible for its stakeholders to enjoy Westin's four states of
  privacy, whenever they are desirable and feasible?
\item Are there some ``privacy patterns'' which could be added to the
  ``design patterns'' of the Gamma et al. \cite{GangOfFour94}, so
  that the architectural foundations for stakeholder privacy can be
  introduced at an early stage of system design?
\item Are there any obvious gaps in the coverage of our initial set of
  privacy patterns?
\end{itemize}

We assume our reader has some prior knowledge of object-oriented design
and the Universal Modelling Language (UML) \cite{UML,Fowler04,Larman04}.

In Section 2, we present our initial set of privacy patterns as UML
class diagrams.  These patterns provide an adequate foundation for the
privacy-aware management of personal identities.

In Section 3, we present subclasses of Entity, allowing us to distinguish
between natural persons (who have privacy requirements) and other actors,
such as computer systems, who do not have privacy requirements.

In Section 4, we present patterns for three contexts in which privacy
requirements may arise for stakeholders.
\begin{itemize}
\item  In the Isolated context, a stakeholder is in Solitude.
\item  In the Secluded context, stakeholders are in Intimacy.
\item In the Public Sphere, stakeholders may enjoy Anonymity, and
are expected to fulfill the social contract of Reserve.
\end{itemize}

\begin{figure}[!t]
\centering
\includegraphics[width=\linewidth]{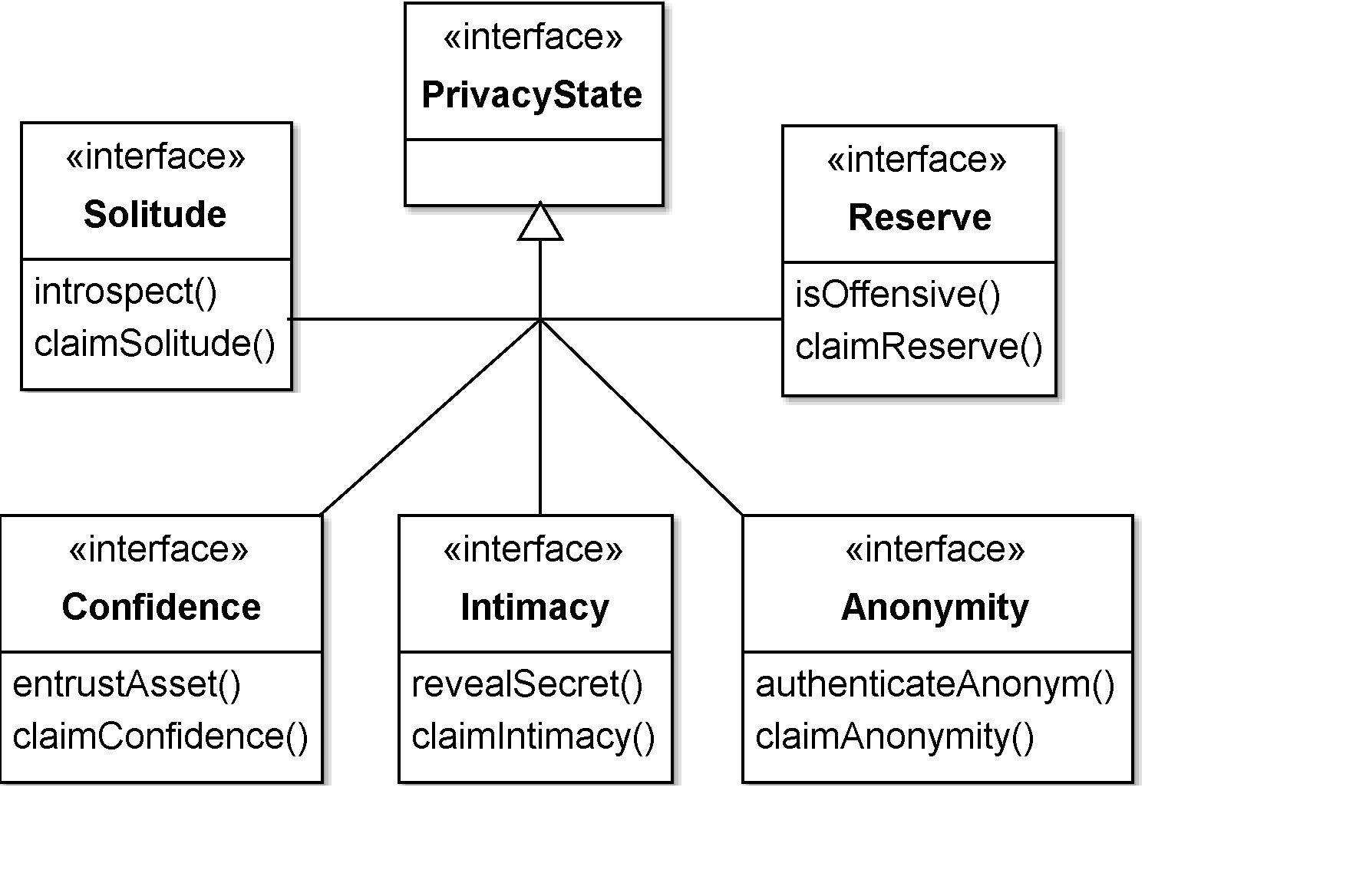}
\caption{Five states of privacy, adapted from Westin \cite{Westin67}}
\label{fig:states}
\end{figure}

In Section 5, we present a complicated context, which we call
Trusting.  Formally, the Trusting context may be considered an
elaboration of the Secluded context defined in Section 4.  However the
Trusting context has so many additional elements, and is so commonly
invoked in legal discussions of informational privacy, that we have
defined Confidence as a fifth subclass of the PrivacyState.  See
Figure~\ref{fig:states}.

\section{Foundations of Structural Privacy}

\begin{figure}[!t]
\centering
\includegraphics[width=\linewidth]{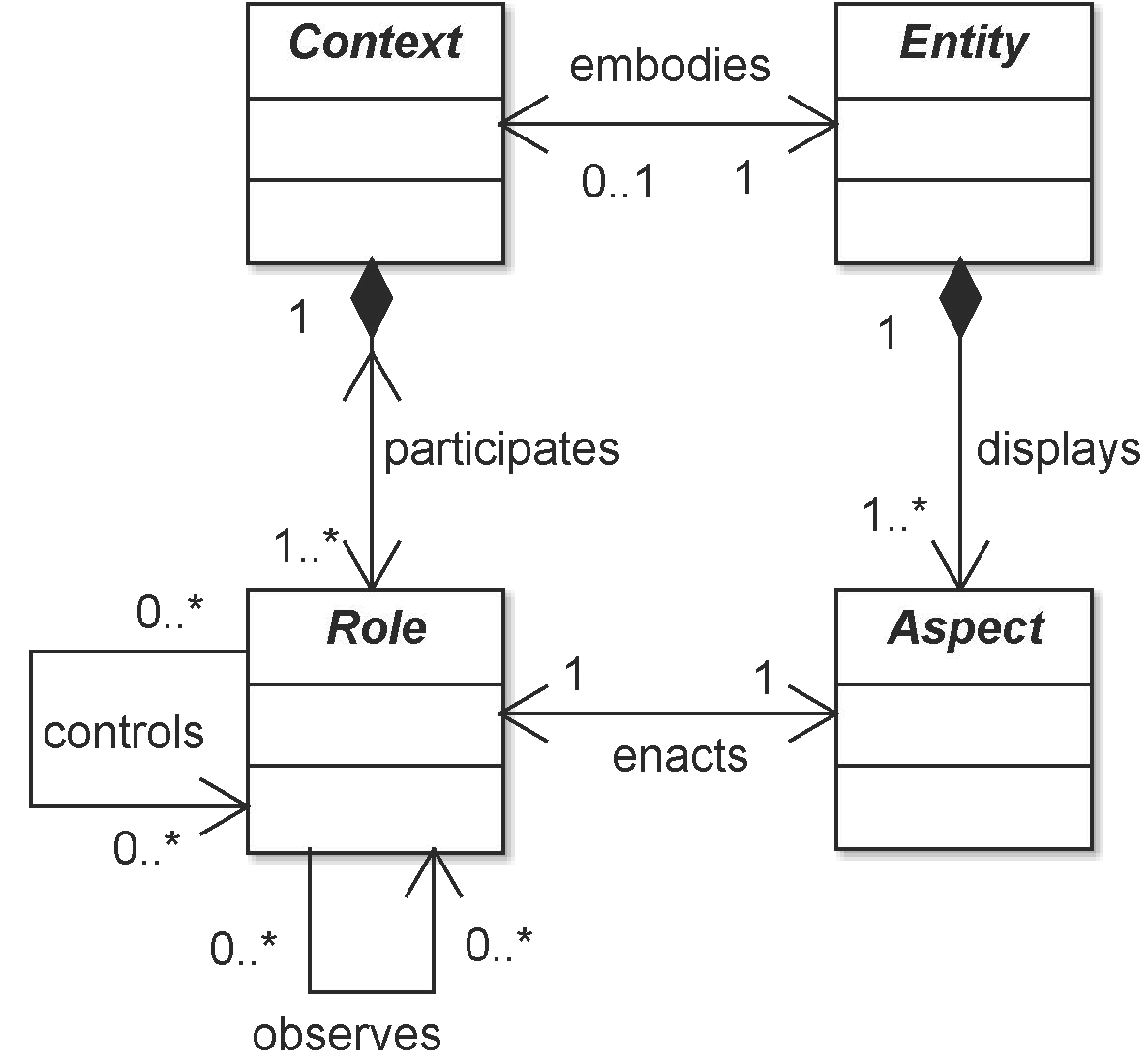}
\caption{An entity displays an aspect, whenever it enacts a role in a
  context.  Contextualised roles may have powers to control and
  observe the activities of other participants in a context.
  Collectively, a context may be embodied as an entity which displays
  an aspect in order to enact a role in some higher-order context.}
\label{fig:crae}
\end{figure}

In this section, we describe a class structure which supports a
contextualised identity for every identifiable entity.  The four
classes of Figure~\ref{fig:crae}, and their four associations, embody
our first four privacy patterns.  We describe each of these patterns
in turn below.  These first four patterns are foundational: all of
our other privacy patterns are elaborations, that is, subclasses, of
classes and associations in Figure~\ref{fig:crae}.

Please note that all four of the classes in Figure~\ref{fig:crae} are
abstract, as indicated by the italic font in their name.  Concrete
subclasses of these classes are defined in later sections of this
article.

\begin{ppat} \emph{Entity-aspect separation.} Entities such as humans
  or computerised systems display one or more Aspects.  The identifier
  for an Aspect does not immediately reveal the identity of the Entity
  who displays this aspect.  Every Aspect must be displayed by an
  Entity; a system should not allow ``orphan'' Aspects.  
\end{ppat}

The structural constraint against orphanage is indicated by the
filled-in diamond on the ``displays'' association between an Entity
and an Aspect in Figure~\ref{fig:crae}.

If an instance Alice of Entity is human, then Alice may display one
set of characteristics and behaviours when she is at work, another set
of characteristics and behaviours in her social life, and a third set
of characteristics and behaviours in her family life.  For analytic
purposes, we may reify these three sets as ``aspects of Alice''.
Alice may herself be such an analyst, and may be consciously
separating her contact-lists and login identities into these three
subsets.

This design pattern, if embodied in every computer system that is used
by Alice, will ensure that the knowledge of any one of her work-login
identities will not provide an immediate link to any of her
extremely-private characteristics and behaviours (i.e. those which are
not displayed in any of her aspects), nor to any of the
characteristics or behaviours which Alice displays to her family and
friends.

We consider this pattern to be a fundament of privacy-aware system
design.  We're not alone in this belief.  Cameron's laws of identity
refer to a ``unidirectional identifier''~\cite{Cameron07}.  The
Jericho Forum's first identity commandment contains the requirement
that ``Core identifiers must only be connected to a persona via a
one-way linkage (one-way trust).''~\cite{JerichoIdentity11} Please
note that there are some terminological differences in our expression
of this design pattern.  Where we write ``Entity'', the commandment
uses the phrase ``core identifier''.  Where we write ``Aspect'', the
commandment uses the word ``persona''.

We suggest that the word ``persona'' should be used only when
referring to an Aspect of a subclass NaturalPerson of Entity, as will
be defined in the next section.  Our reasoning is that the privacy
rights and expectations of a human (a.k.a. a natural person, in legal
discourse) may be directly supported by this design pattern in its
subcase of an entity-persona separation, and they may be indirectly
supported by this design pattern in its general case of an
entity-aspect separation.  Using a different name for the Persona
subclass of an Aspect would, we suggest, allow requirement analysts
and systems architects to communicate more accurately and succinctly.

In Latin, \emph{persona} is an actor's mask, connoting that it is worn
by a human when playing a role in a play.  A play would be represented
as an instance of a Context in our class system.  If an actor were a
completely-computerised entity, we think it would be most appropriate
to use an inhuman word such as ``interface'' rather than ``persona''
when speaking of its aspects.

Linguistically: all of the classes in our patterns are concrete nouns.
The relations between classes are verbs.  Interfaces are abstract
nouns or occasionally adjectives.  Our class diagrams thus define a
specialised ontology for the discussion of privacy in complex systems.

Readers who are not familiar with the concept of a design pattern may
imagine that it is a prescriptive form of architectural description
\emph{i.e.} specifying absolute requirements on its structural design.
However a design pattern is not prescriptive, instead it is a
trope or commonly-used design motif.

\begin{quote}
  Design patterns capture solutions that have developed and evolved
  over time.  Hence they aren't the designs people tend to generate
  initially.  They reflect untold redesign and recoding as developers
  have struggled for greater reuse and flexibility in their software.
  Deisgn patterns capture these solutions in a succinct and easily
  applied form.~\cite{GangOfFour94}
\end{quote}

We now return to our discussion of Figure~\ref{fig:crae}, focussing on
the association between an Aspect and a Role.

\begin{ppat} \emph{Aspect-role separation.} An Aspect is influential
  only when it is enacting a Role in some Context.  The Role puts
  constraints and expectations on the range of activity of an Aspect.
  When an Entity (through an Aspect) is playing a Role, it may be able
  to observe and control the activities of other Role-enacting and
  Aspect-displaying Entities.  Depending on the Context and the Role,
  it may be possible for a Role-enacting Aspect to be identified.
  Such identifications are crucially important when establishing
  accountabilities for this Aspect in other Contexts, e.g. in a
  judicial proceeding.
\end{ppat}

The aspect-role-separation pattern allows us to express the social and
legal determinants of privacy, in a context-dependent way.  In Section
4, we develop important subclasses of Context, for example Isolated,
in which there are stereotyped Roles such as the Isolate and the
Intruder.  Each Role is played by an Aspect which implements the
interface (e.g. Solitude) relevant to this role in this context.  The
interface defines the stereotypical behaviours (such as
\verb|introspect()|) which are socially or legally expected of any
Aspect enacting this particular subclass of Role.

In a UML class diagram, an arc indicates that an instance of a class
is able to access the data and methods of another instance.  We use
labels to indicate the type of access.  In particular, an arc labelled
``controls'' indicates that a method could be invoked, or a data
field could be written.  An arc labelled ``observes'' indicates
that a data field could be read.  This is informal semantics.  To
formalise these semantics, we could subtype an association class.
However we see very little benefit in expressing this subtyping
formally, especially since typed associations are rarely seen in UML
class diagrams for software systems.

\begin{ppat} \emph{Role-context separation.} Every Role is
  participating in a Context.  Orphaned Role objects should be
  garbage-collected, because a role is ineffectual if it is not
  associated with any context.
\end{ppat}

The reader may find it helpful to think of the list of Roles in a
Context as being analogous to the Dramatis Personae of a play.  One of
the tasks of the director of a play is to ensure that, whenever it is
performed, all of its currently-active roles are being enacted by some
actor.  It is possible for the same actor to play multiple roles in
the same play, perhaps even simultaneously.  However our first design
pattern suggests that a privacy-aware system will prevent any entity
from playing a role in any context as itself; instead, every entity is
required to put on some mask (an Aspect) whenever they are on stage.
This mask will prevent them from revealing information (such as the
actual colour of their skin) which would inappropriate in the context
of this play.  The aspect of each actor must of course conform to the
role they are playing.  It should not be possible for anything to
happen on the stage, unless it is being performed by some Role defined
in this play.

\begin{ppat} \emph{Context-entity separation.} Every Context is the
  embodiment of some Entity.  Some Entities embody a Context.
\end{ppat}

The ``embodies'' association closes the loop of our foundational
privacy patterns, by allowing conglomerates or corporations (as
structured within a Context) to behave as entities in some
higher-level context.  For example, a corporation may have many
employees and a complex decision-making apparatus.  However from a
legal perspective, a corporation is answerable to a judicial authority
in the context of a legal proceedings.

\begin{figure}[!t]
\centering
\includegraphics[width=\linewidth]{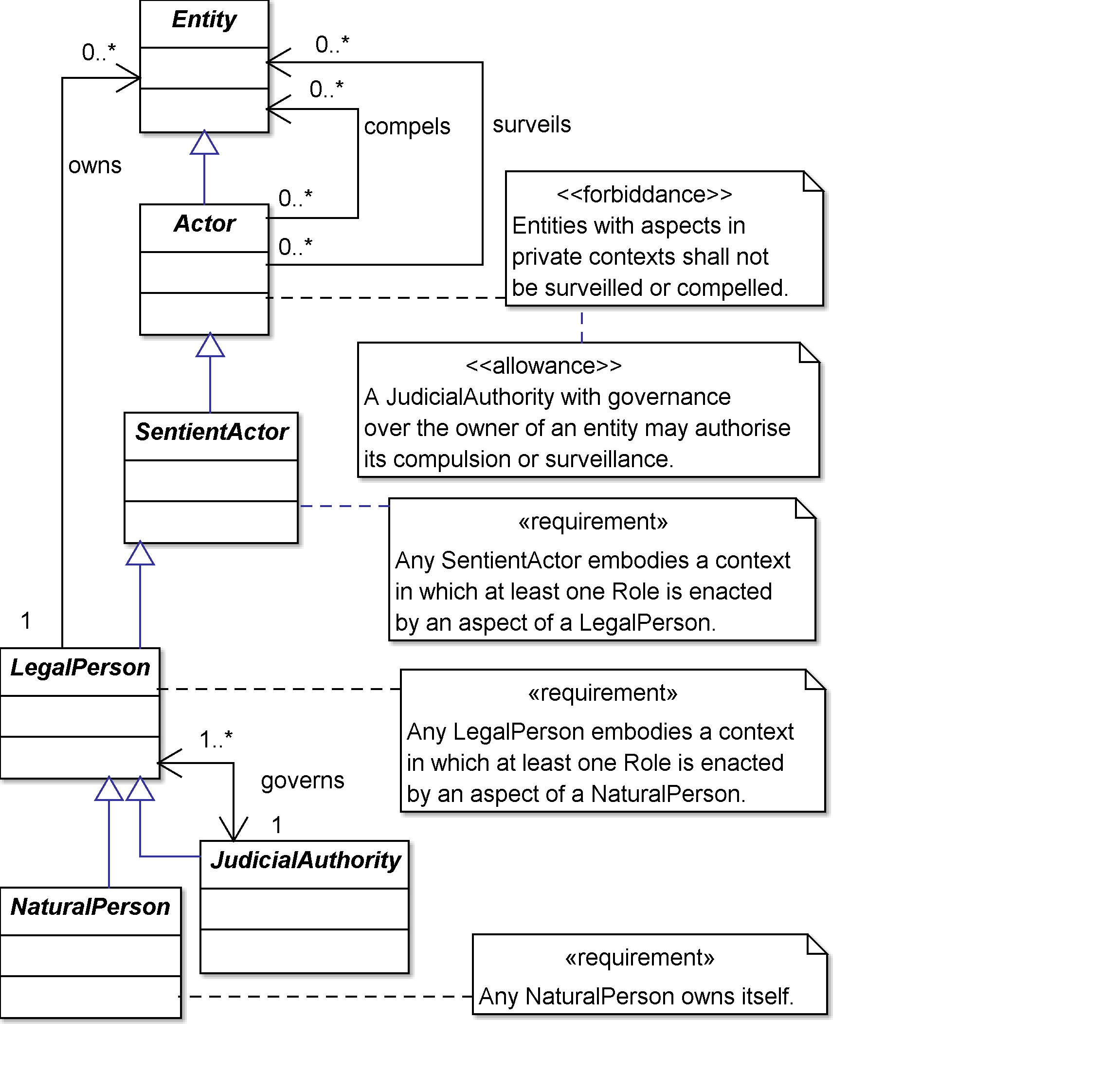}
\caption{Five subtypes of Entity, with relations of ownership,
  governance, surveillance and compulsion.  We model a human being as
  an instance of NaturalPerson.  An informally-constituted group of
  people is a SentientActor.  If a group of people is incorporated as
  a legal entity, then it is a LegalPerson.}
\label{fig:entity}
\end{figure}

\section{Types of Entity}

We move now to Figure~\ref{fig:entity}, which defines five subtypes of
Entity.  This proliferation of entity types may seem unfortunate,
however we cannot see how our privacy patterns could have fewer
abstract classes inheriting from Entity, while still distinguishing
privacy requirements from other types of requirements.  In
Figure~\ref{fig:entity}, the most general class (Actor) is drawn at
the top, and the most specific classes are at the bottom.  Please
recall that in the UML, an open-ended triangle indicates the direction
of the inheritance relation between two classes.

An Actor is a specialised form of Entity: an actor may surveil or
compel other entities, whereas any class which inherits directly from
an Entity can neither suveil nor compel.  A document is an example of
an entity which is not able to surveil or compel.  A computer is an
example of an entity that is able to surveil or compel.  We have
introduced surveillance and compulsion into our patterns to indicate

\begin{itemize}
\item how a privacy requirement could be breached, by any actor who
  uses surveillance or compulsion to subvert the one-way entity-aspect
  relation, and also
\item how a privacy requirement could be enforced, by any actor which
  is trusted to use its surveillances and compulsions only to detect
  privacy breaches, and to respond appropriately to any detected (or
  alleged) breaches.
\end{itemize}

We use callouts to express requirements, as in the SysML
profile~\cite{SysML} of UML~\cite{UML}.  Callouts are linked to a
class, or to another callout, by a dashed line.  We use five
stereotypes in our annotations: \guillemotleft
requirement\guillemotright, \guillemotleft forbiddance\guillemotright,
\guillemotleft allowance\guillemotright, \guillemotleft
obligation\guillemotright, and \guillemotleft
exemption\guillemotright.

A \guillemotleft requirement\guillemotright\ is a correctness
constraint, i.e. a property of a system which could be formally
verified.  For example, the anti-slavery requirement at the bottom of
Figure~\ref{fig:entity} is, formally, a constraint on the ``owns''
relation of every instance of NaturalPerson.

We use our other four stereotypes as a rough classifier of privacy
requirements.  A requirement that curtails the range of acceptable
actions by an actor is a \guillemotleft forbiddance\guillemotright;
and its text field must specify which action(s) SHALL NOT be performed
by this actor.  A forbiddance may be derogated by an attached
\guillemotleft allowance\guillemotright\ which specifies some
action(s) which MAY be performed.  Alternatively, a privacy
requirement may be expressed as an allowance which is derogated by one
or more forbiddances.

Another common features we have observed, when reviewing the
privacy-structures of existing systems for this article, is that a
privacy requirement curtails the range of inactions by an actor.  We
use the \guillemotleft obligations\guillemotright\ stereotype on such
requirements, and the accompanying text field must specify which
action(s) SHALL be performed.  Derogations on obligations are called
\guillemotleft exemptions\guillemotright, and these specify actions
which MAY NOT be performed.

For logical completeness, we will allow a privacy requirement to be
expressed as an exemption which is derogated by some obligation(s),
although to date we have not elicited any requirement that would be
naturally expressed in this format.

Figure~\ref{fig:entity} exhibits a privacy pattern of a judicial
authority with a governance relation over the entities it recognises
as legal persons.  A privacy-sensitive judical authority will prohibit
surveillance and compulsion, except in cases where these powers are
necessary to its dispensation of justice.

\begin{ppat} \emph{Surveillance and compulsion are forbidden.}
  Actors are generally forbidden from surveilling and compelling other
  entities, however there is an important exception to this
  forbiddance.  A judicial authority may allow some surveillance and
  compulsion.
\end{ppat}

Figure~\ref{fig:entity} contains a second privacy pattern: the concept
of legal ownership, with a structural restriction against
slavery.  We note, in passing, that it may be possible to develop
design patterns for all generally-accepted human rights.

\begin{ppat} \emph{Entities are owned.}  Any legal person may own
  any entity, unless the second entity is another natural person.  The
  legal ownership of an entity implies a legally-enforceable right to
  compel, surveil, control and observe that entity.  Every entity has
  an owner.
\end{ppat}

We note that the ``owns'' relation of Figure~\ref{fig:entity} is 1-way
navigable.  This models the possibility that an entity, such as a
document, does not carry a record of its current owner.  In particular
systems, it might be appropriate to specify a navigation from an
entity to its owner as an \guillemotleft allowance\guillemotright, or
as a \guillemotleft forbiddance\guillemotright, if obtaining
information about the ownership of this entity is privacy-sensitive.

We note that different stakeholders will, in general, have different
privacy requirements.  During the privacy-requirements elicitation for
a system, these differences could be recorded in stakeholder-specific
callouts on a class diagram for this system.

We also note that our privacy patterns will support many theologies.
In particular, an atheist may define a LegalPerson that has null (or
randomly-generated) method bodies; this entity could serve as the
owner of all entities that are not subject to any other judicial
authority.  By contrast, a monotheist may construct a class hierarchy
in which God is a LegalPerson which governs itself, and which has a
JudicialAuthority over all other JudicialAuthorities.  Because privacy
patterns are suggestive rather than prescriptive, the architects of a
system may develop a class hierarchy in which some entities do not
have explicitly-represented owners.

Finally, we note that artificial persons such as corporations may make
claims to privacy requirements~\cite{Pollman2014}, as may
informally-constituted groups such as clusters of friends,
participants in religious ceremonies, and extended families.  In such
cases, the class analyst should construct an entity to embody the
claimant's group, then attempt to map their group-privacy claims onto
the privacy patterns of the next section.  If this is infeasible, we
hope the analyst will contact us with a description of the group's
unrepresentable privacy claim, or with any new privacy pattern they
devise for its representation.

\section{States of Privacy}

We move now to Figure~\ref{fig:states}. The five interfaces in this
figure define five distinct states or modalities of privacy.  Four of
these states were identified by Westin in his 1967 survey of privacy,
as understood by contemporary ``anthropologists, psychologists,
biologists, physicists, historians, and psychiatrists, as well as
philosophers, lawyers, and laymen''~\cite{Westin67}.  Westin argues
that the states of solitude, intimacy, reserve, and anonymity are
recognised and protected in a wide range of cultures.

In our reductionist treatment of Westin's conception of privacy, we
assert that any state of privacy may be claimed by any person, or by
any group of people, in any context.  We consider it the job of the
requirements analyst to discover the particular forms of privacy that
are likely to be claimed by stakeholders, and to determine which of
these privacy claims can be accepted or rejected automatically.  We
consider it the job of the systems implementer to provide affordances
which, ideally, will make it easy (or even automatic) for users
\begin{itemize}
\item to make and withdraw claims to privacy,
\item to enquire into the status of their outstanding claims,
\item to resolve problematic claims as fairly and quickly as possible,
  with recourse to appropriate external parties for the most
  problematic claims, and
\item to effectively defend privacy claims, almost always without
  recourse to external enforcement agents, but sometimes by engaging
  other users of the system who have a vested interest in this claim
  or this type of claim.
\end{itemize}

\begin{figure}[!t]
\centering
\includegraphics[width=\linewidth]{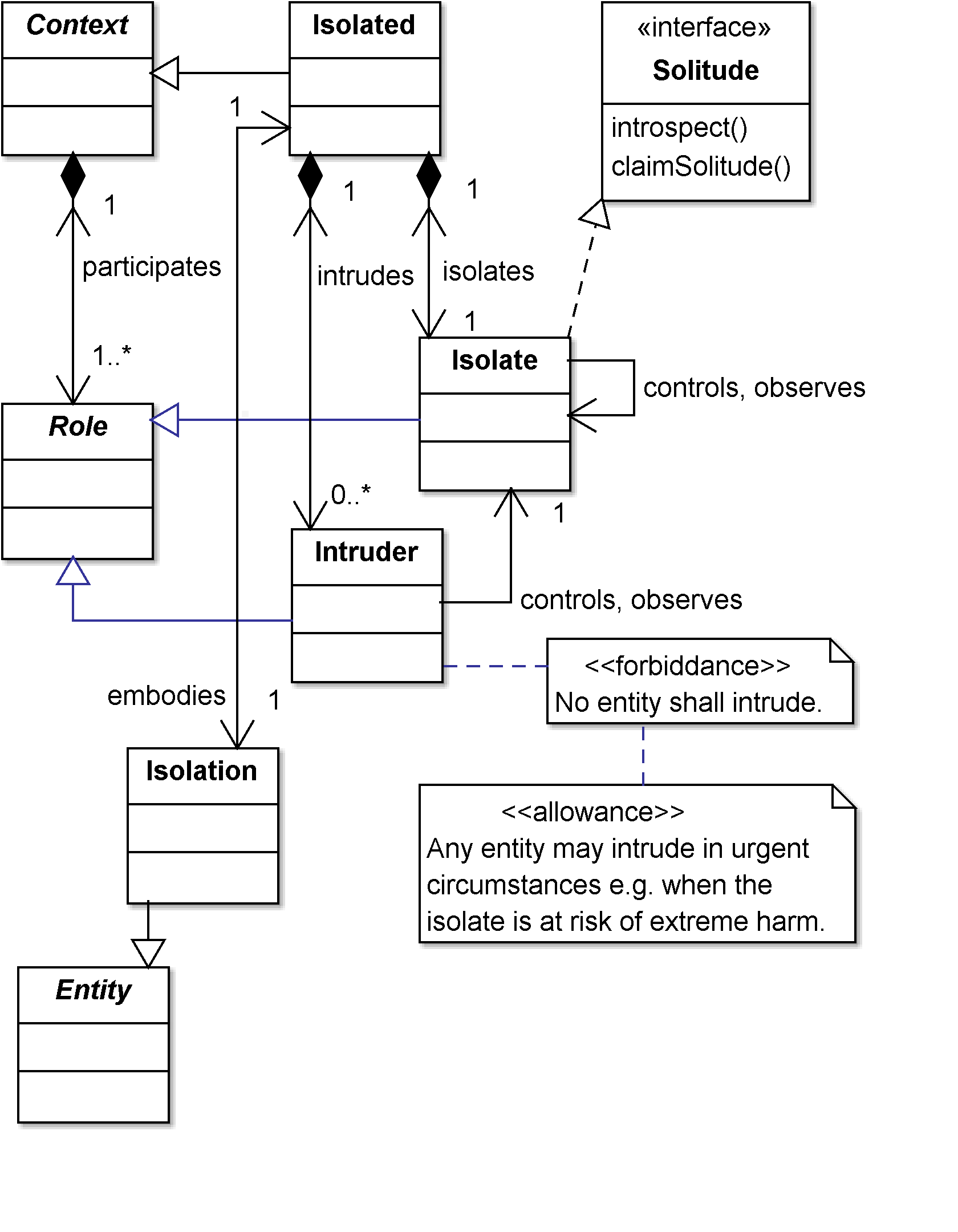}
\caption{Privacy Pattern 7: Solitude.  An entity with an aspect
  playing the Isolate role is able to introspect, but is unable to
  control or observe any other entity.  Generally, other entities are
  forbidden from intruding on an isolate.  Urgent intrusions may be
  allowed by the governor of the isolate, or by the societies to which
  the isolate belongs.}
\label{fig:solitude}
\end{figure}

\subsection{Solitude}

Figure~\ref{fig:solitude} is a design pattern for a context which
realises the Solitude state of privacy.  We elicited this pattern from
the following passage in Westin's monograph \cite{Westin67}:

\begin{quote}
  The first state of privacy is solitude; here the individual is
  separated from the group and freed from the observation of other
  persons. He may be subjected to jarring physical stimuli, such as
  noise, odors, and vibrations. His peace of mind may continue to be
  disturbed by physical sensations of heat, cold, itching, and
  pain. He may believe that he is being observed by God or some
  supernatural force, or fear that some authority is secretly watching
  him. Finally, in solitude he will be especially subject to that
  familiar dialogue with the mind or conscience. But, despite all
  these physical or psychological intrusions, solitude is the most
  complete state of privacy that individuals can achieve.
\end{quote}

The Isolated context in Figure~\ref{fig:solitude} is embodied in a
class of entity which we have named Isolation.  This embodiment allows
an aspect of an instance of Isolation to be persistently identifiable
while it is playing a role in a surrounding context.  For example, if
the surrounding context were in an airport, it could have a guarantor
of solitude who responds to claimSolitude() calls from nearby isolates,
by offering them a rental contract for a currently-vacant sleeping
pod.

\begin{figure}[!t]
\centering
\includegraphics[width=\linewidth]{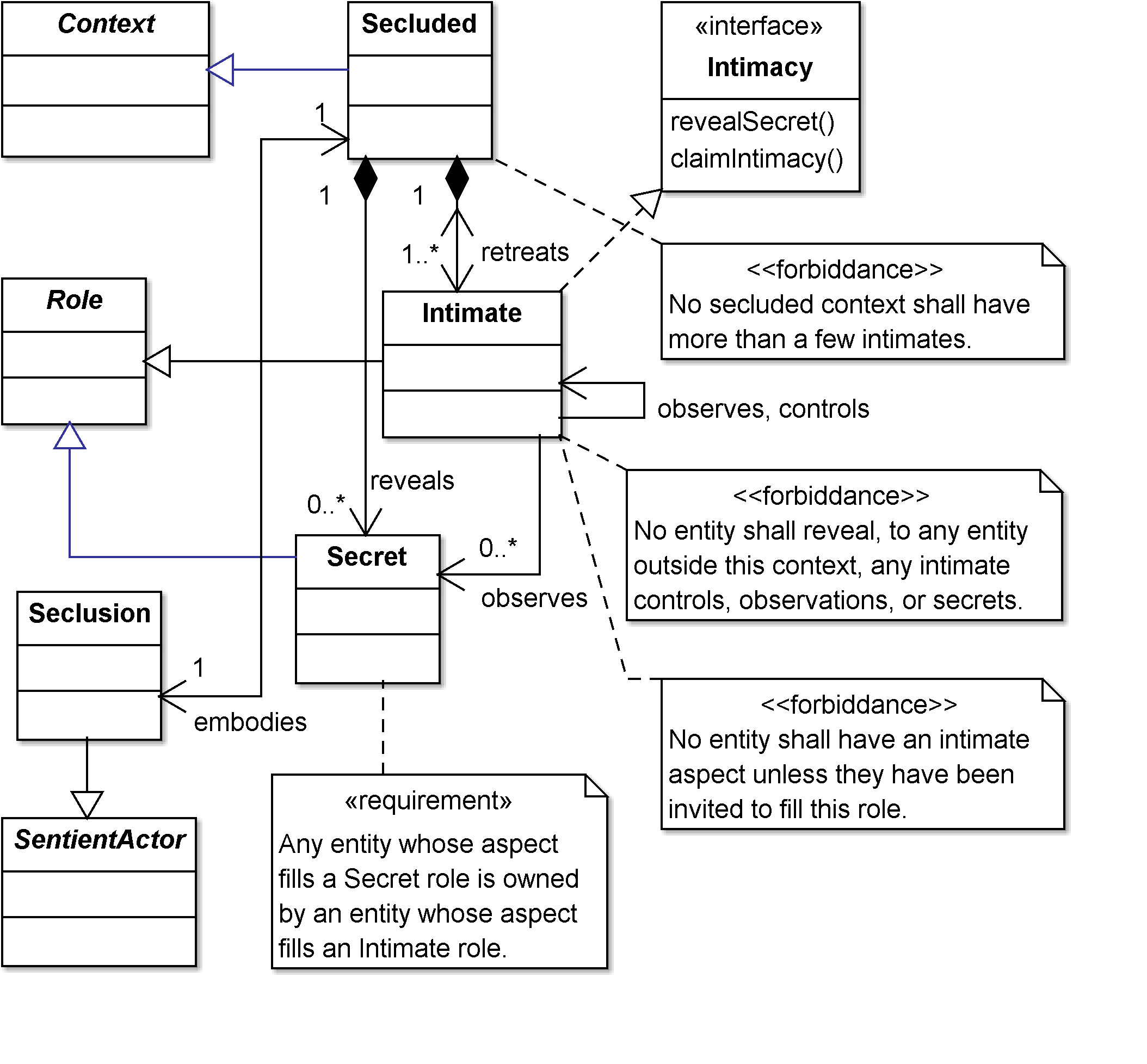}
\caption{Privacy Pattern 8: Intimates share secrets in a secluded context.}
\label{fig:intimacy}
\end{figure}

\subsection{Intimacy}

We elicit the Secluded context of Figure~\ref{fig:intimacy} from
the following passage in Westin~\cite{Westin67}:

\begin{quote}
  In the second state of privacy, intimacy, the individual is acting
  as part of a small unit that claims and is allowed to exercise
  corporate seclusion so that it may achieve a close, relaxed, and
  frank relationship between two or more individuals. Typical units of
  intimacy are husband and wife, the family, a friendship circle, or a
  work clique. Whether close contact brings relaxed relations or
  abrasive hostility depends on the personal interaction of the
  members, but without intimacy a basic need of human contact would
  not be met.
\end{quote}

We derive two privacy requirements on intimates: they must not join an
existing intimate group without an invitation, and they must not
reveal anything about the intimacy to any outside party.  We add a
structural requirement, that the Secret role must be filled by an
aspect of an entity that is owned by one of the intimates. 

Westin's passage suggests that an additional requirement may be
commonly elicited: that there be some upper limit on the size of an
intimate group.  This would limit the risk that claims to seclusion
are used to promote a political cause or to serve a commercial
purpose, rather than for a ``close, relaxed, and frank relationship''
between a few individuals.

\begin{figure*}[!t]
\centering
\includegraphics[width=.8\linewidth]{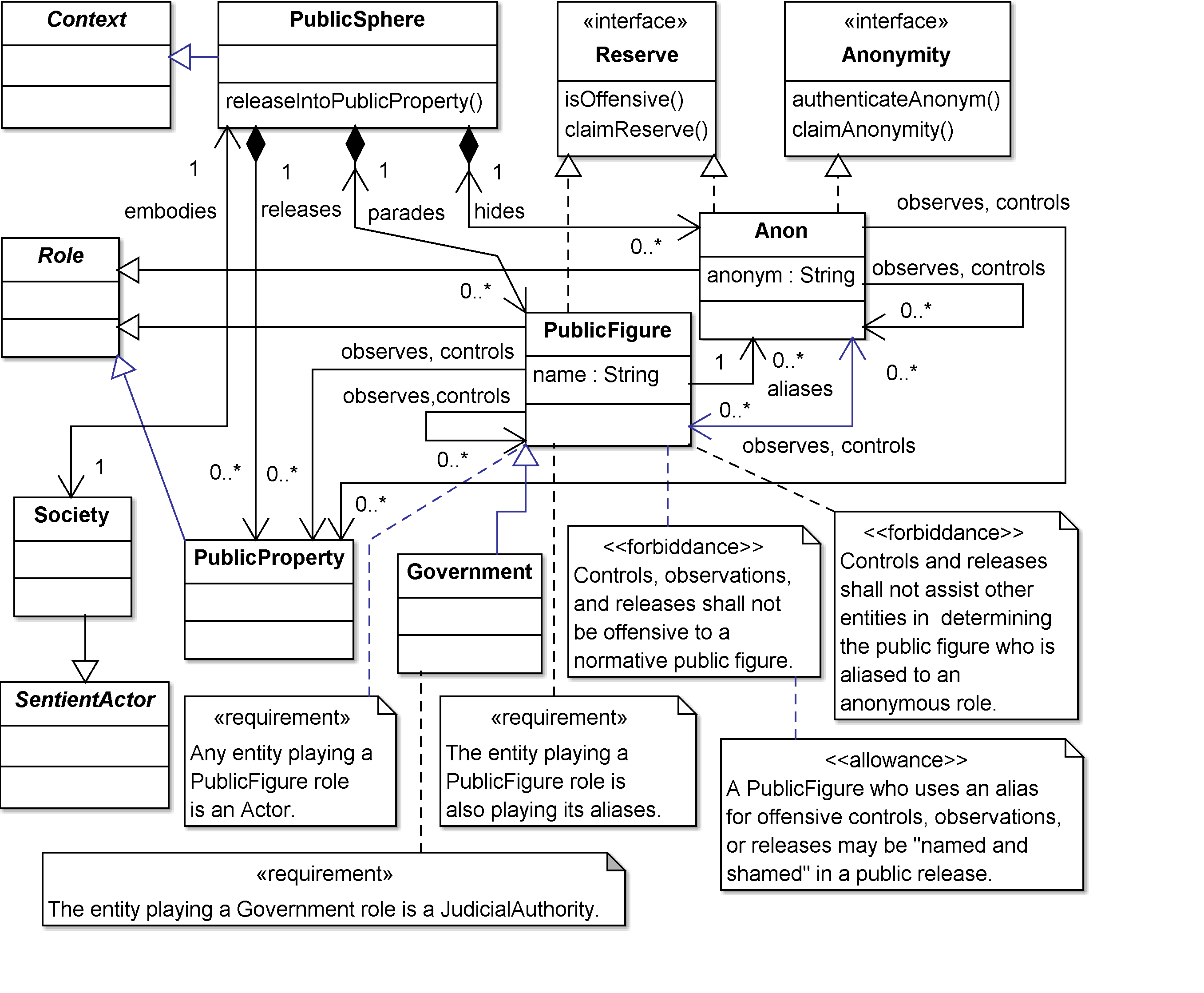}
\caption{Privacy Pattern 9: A public sphere with Anonymity and
  Reserve.  The members of a public sphere, collectively, are embodied
  as a society which self-regulates its conceptions of anonymity and
  reserve, and which administers all public property.  Societal
  sanctions include ``naming and shaming''.  Individuals may appeal to
  their governor, if they feel wrongly treated by their society.}
\label{fig:publicsphere}
\end{figure*}

\subsection{Anonymity and Reserve} Figure~\ref{fig:publicsphere} is
our most complex schema.  In it, we have modelled the states of
Reserve and Anonymity in a context (PublicSphere) which is embodied as
a Society.  There are three types of actor in our PublicSphere:
PublicFigures (who have names which could be replicated in other
PublicSphers), anonymous roles (each of which is aliased to exactly
one public figure), and instances of PublicProperty (which model
newspapers, advertisements, and any other identifiable entities which
are displaying observable and controllable aspects in the commons of
this society).

The privacy requirements in this schema are restrictions on the
behaviour of public figures.  They are expected, by their society,
to regulate their own behaviour -- and that of any anonymous
aliases they may be using.  We elicited the state of Anonymity from
the following passage in Westin's monograph~\cite{Westin67}.

\begin{quote}
  The third state of privacy, anonymity, occurs when the individual is
  in public places or performing public acts but still seeks, and
  finds, freedom from identification and surveillance. He may be
  riding a subway, attending a ball game, or walking the streets; he
  is among people and knows that he is being observed; but unless he
  is a well-known celebrity, he does not expect to be personally
  identified and held to the full rules of behavior and role that
  would operate if he were known to those observing him. In this state
  the individual is able to merge into the ``situational landscape.''
  Knowledge or fear that one is under systematic observation in public
  places destroys the sense of relaxation and freedom that men seek in
  open spaces and public arenas.

  \quad Still another kind of anonymity is the publication of ideas
  anonymously.  Here the individual wants to present some idea
  publicly to the community or to a segment of it, but does not want
  to be universally identified at once as the author-especially not by
  the authorities, who may be forced to take action if they ``know'' the
  perpetrator. The core of each of these types of anonymous action is
  the desire of individuals for times of ``public privacy.''
\end{quote}

We elicited the state of Reserve from the following passage.

\begin{quote}
  Reserve, the fourth and most subtle state of privacy, is the
  creation of a psychological barrier against unwanted intrusion; this
  occurs when the individual's need to limit communication about
  himself is protected by the willing discretion of those surrounding
  him. Most of our lives are spent not in solitude or anonymity but in
  situations of intimacy and in group settings where we are known to
  others. Even in the most intimate relations, communication of self
  to others is always incomplete and is based on the need to hold back
  some parts of one's self as either too personal and sacred or too
  shameful and profane to express. This circumstance gives rise to
  what Simmel called ``reciprocal reserve and indifference,'' the
  relation that creates ``mental distance'' to protect the
  personality. This creation of mental distance -- a variant of the
  concept of ``social distance'' -- takes place in every sort of
  relationship under rules of social etiquette; it expresses the
  individual's choice to withhold or disclose information -- the
  choice that is the dynamic aspect of privacy in daily interpersonal
  relations. Simmel identified this tension within the individual as
  being between ``self-revelation and self-restraint'' and, within
  society, between ``trespass and discretion.'' The manner in which
  individuals claim reserve and the extent to which it is respected or
  disregarded by others is at the heart of securing meaningful privacy
  in the crowded, organization-dominated settings of modern industrial
  society and urban life, and varies considerably from culture to
  culture~\cite{Westin67}.
\end{quote}

The boundaries of a reserve may be changed, from time to time, by the
active members of that society.  This implies that elicitations of a
reserve should be revalidated, and possibly revised, as the society
evolves.

As with most other types of privacy requirements, we do not expect the
requirements associated with reserves to be enforced accurately by a
computerised system.  However, it would be feasible to design a system
in which a computerised actor is a PublicFigure in a society, and to
advertise (via a release to PublicProperty) the availability of its
isOffensive() method.  If the members of the society come to a general
agreement that this method has few false-positives, then it could be
reliably used as a detector of some forms of offensive behaviour in
that society.  If it is feasible to develop a socially-acceptable
automated response mechanism for automatically-detected violations,
then the social reserve is being technically enforced.  However in
any case where the automated responses are socially inappropriate,
we would say that the social reserve is being violated by this
computerised system -- and we would expect a corrective response via
social, legal, or economic pressure on the owner of the system.

The governmental role in the public sphere is played by a
JudicialAuthority, as originally defined in Figure~\ref{fig:entity}
and as referenced in the bottom callout of Figure~\ref{fig:publicsphere}.

The two forbiddances in Figure~\ref{fig:publicsphere} are illustrative
but not prescriptive.  The first is a general statement of reserve,
framed in a manner that suggests a common-law approach to its
definition whereby the isOffensive() method of a normative public
figure such as a judge or kaumatua (a revered elder, in Maori) is used
to determine whether or not a sanction is appropriate.  The second is
supportive of anonymity, in the sense that actors are forbidden from
assisting each other in piercing the veil of anonymity.  The allowance
in this figure is intended to suggest the social conventions which
distinguish an unoffensive publication from an actionable breach of a
generally-acknowledged reserve.

We have not specified, in this figure, whether the anonymity is
linkable, that is, whether the binding of an aspect to an Anon
instance is persistent (so that an Anon instance has a pseudonym that
is linked to a reputational history), or (in an extreme form of
unlinkability) each Anon instance acts at most once and its enacting
entity is careful not to disclose any information about their prior
anonymous actions.  The implementation of the authenticateAnonym()
method would greatly affect linkability, for example if it always
returns false then the Anon instances would be offering no assistance
to anyone who is attempting to determine their identity.  At the other
extreme, an authenticateAnonym() call might return a
cryptographically-sound proof of a claim to an identity that is
specified, in the parameters of the authenticalAnonym() call, as an
instance of an Anon in a PublicSphere. 

\begin{figure*}[!t]
\centering
\includegraphics[width=.8\linewidth]{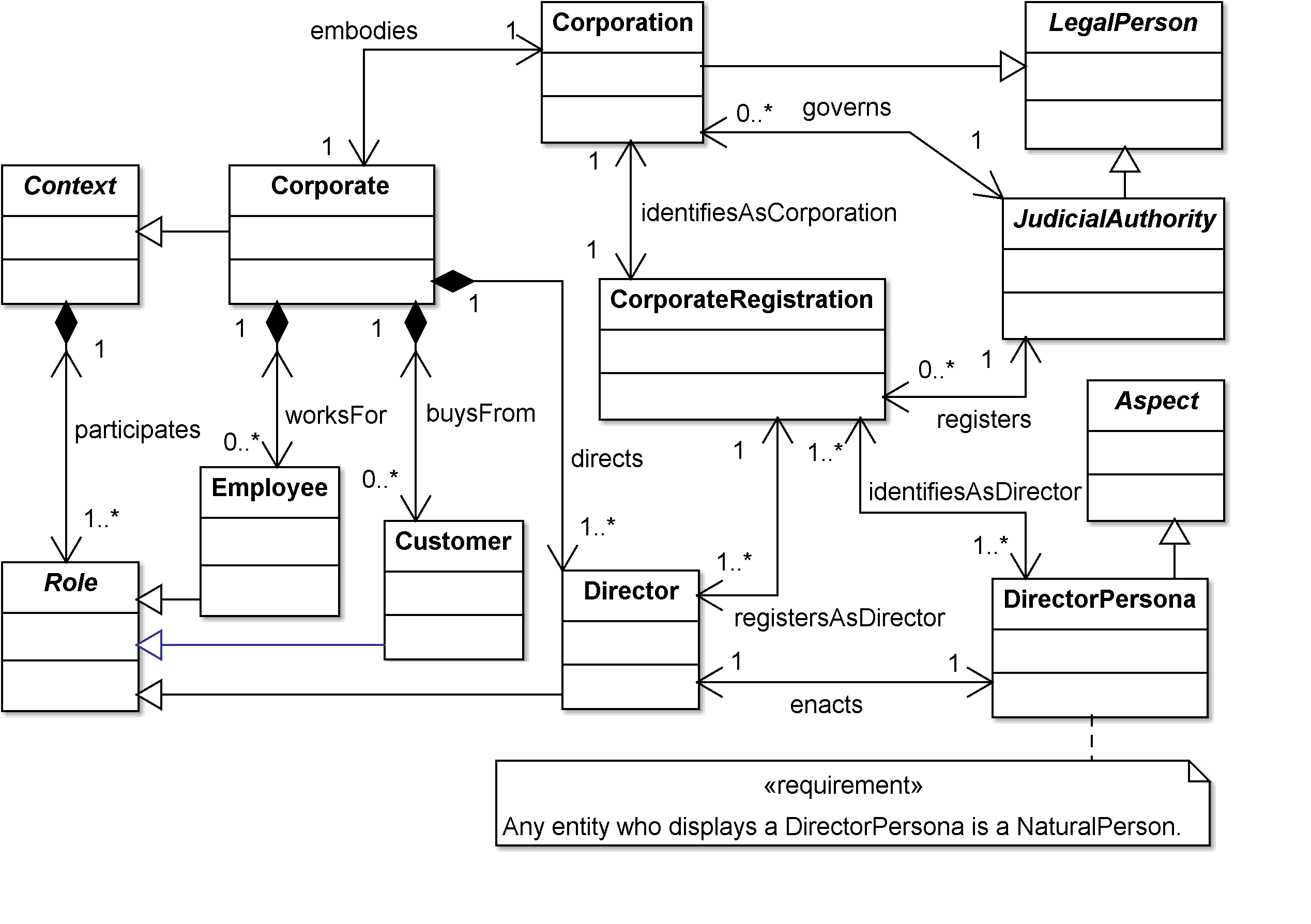}
\caption{A corporation with registered directors.}
\label{fig:corporation}
\end{figure*}

\begin{figure}[!t]
\centering
\includegraphics[width=\linewidth]{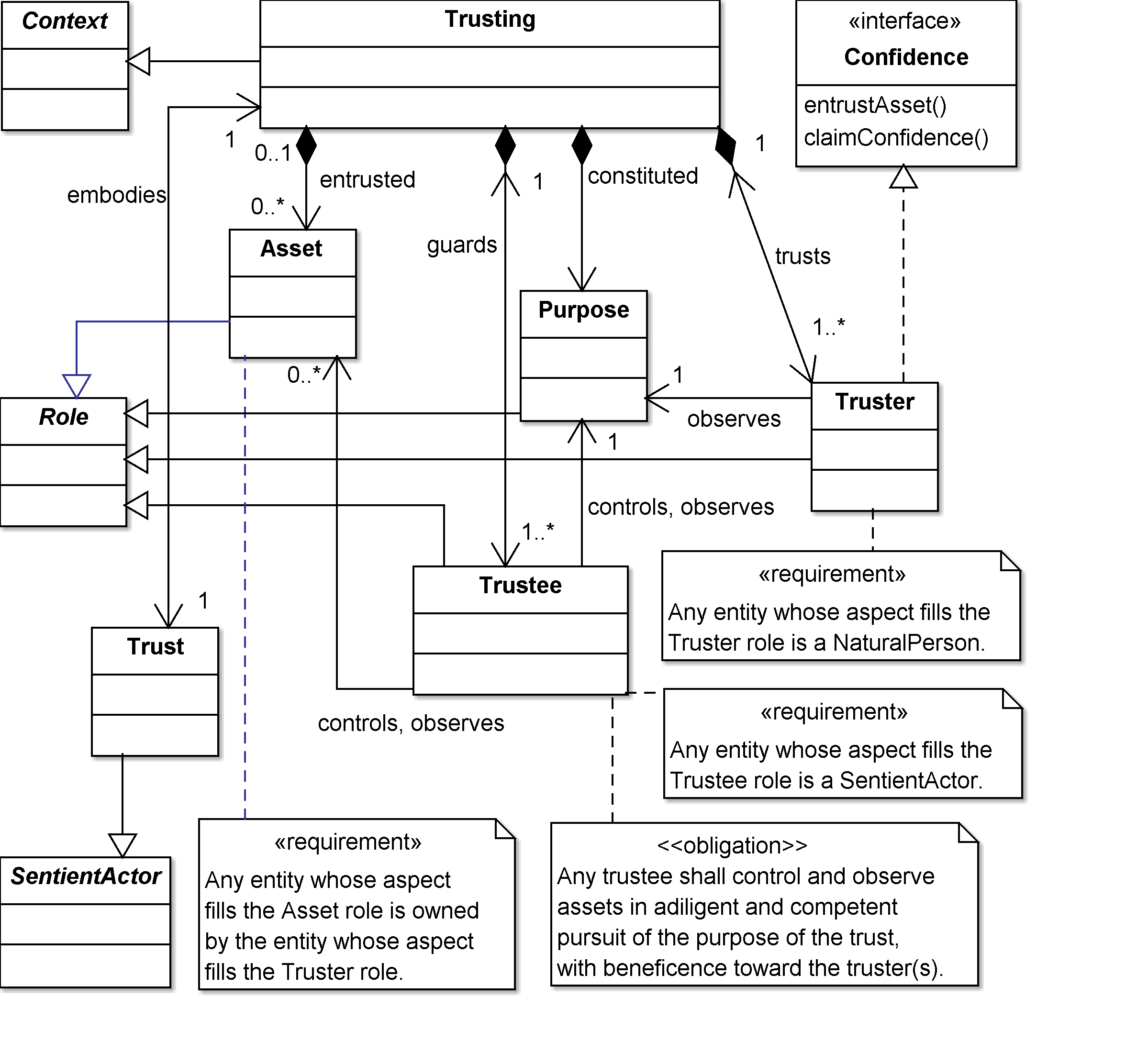}
\caption{Privacy Pattern 10: Confidence.  The state of Confidence is
  structurally similar to the state of Intimacy, however the context
  for Confidence has a very pronounced asymmetry between the Truster
  and the Trustee, whereas all intimates have equivalent powers.  A
  Trustee has a duty of care for the Truster's asset.}
\label{fig:confidence}
\end{figure}

\subsection{Confidence}

We have added ``confidence'' to Westin's list of states.  This fifth
state allows our schemas to directly represent a conception of privacy
which has been adopted by many technologists and policy analysts.  In
this conception of privacy, sometimes called ``information privacy''
\cite{DeCew97}, an individual's control over the confidentiality of
their personal information is considered to be the primary (or even
the only) meaning of the word ``privacy''.

We elicited the Confidence state of privacy from the EU's Data
Protection Directive.  Figure~\ref{fig:confidence} is our reductionist
interpretation of this complex directive.  Each member state of the EU
will comply with this directive in its own way; but generally
speaking, each member state is obligated to impose a duty of care
(i.e.  a set of obligations) on every controller of personal data who
falls within their jurisidction.  The legal situation of a controller
is thus roughly comparable to that of a trustee, in a trust which
holds the personally-identifiable information of the data subjects as
its asset.

We note that our schema for Confidence models an informal trust, not a
legal trust.  A legal trust is a specialisation of our privacy pattern
for Confidence such that the trust is a LegalActor, and not merely a
SentientActor.  Generally, a JudicialAuthority requires all Trustees
of its legally-registered trusts to be LegalActors within its
jurisdiction.

\subsection{Corporation}
In Figure~\ref{fig:confidence} we depict a corporation as an identifiable
entity in our class hierarchy.  We do not elicit any privacy requirements
from this depiction, however we include this diagram in our article 
because corporations are often cast as antagonists or protagonists in
contemporary discourse about privacy.

\section{Discussion}

The dozens of classes and relations defined in this article are
sufficient to support our initial set of privacy patterns.  We
elicited these privacy patterns by analysing texts, such as Westin's
monograph~\cite{Westin67}, that describe various conceptions of
privacy.

We do not imagine that our initial set of privacy patterns is
complete.  Instead, we think that libraries of design patterns must
always be open to addition, as new types of requirements provoke
system architects to create novel solutions which can be generalised
for re-use by other architects.  To this end: we invite the reader to
analyse textual descriptions of privacy requirements, and to conduct
analyses of additional privacy-sensitive systems, in a quest for the
discovery of additional privacy patterns of general relevance.

Finally, we invite the reader to consider whether it might be
helpful, in their context, to extend the UML use-case diagram so that
it is expressive of what we might call ``privacy cases''.  We have
some initial ideas on how this might be done, and would welcome
collaborators in this endeavour.

\section{Acknowledgements} Many aspects of this article were shaped by
our enjoyable and enlightening discussions of identity, access, and
entitlement within The Jericho Forum, from 2005 to 2012.

The figures in this article were prepared with open-source software
ArgoUML~\cite{argoUML}.

\bibliographystyle{IEEEtran}    % or some other suitable package.

\bibliography{IEEEabrv,pprivacy}  % often included from a separate file.

\end{document}